\documentclass[a4paper]{mem}
\usepackage{natbib}
\usepackage{graphicx}
\usepackage[a4paper]{hyperref}
\idline{0}{1}
\begin{document}
   \title{The RGB Bump in dwarf Spheroidal galaxies: \\ 
   discovery and perspectives.}

   \author{M. Bellazzini \inst{1} 
    \fnmsep
    \thanks{The results presented in this contribution are part of a long term
    programme at OAB. The other members of the team are: F.R. Ferraro, 
    L. Origlia, L. Monaco, and E. Pancino.}
}

   \offprints{Michele Bellazzini}
\mail{bellazzini@bo.astro.it}

   \institute{INAF - Osservatorio Astronomico di Bologna, 
via Ranzani 1, 40127 Bologna, Italy \email{bellazzini@bo.astro.it}\\ 
}

   \abstract{I report on the recent discoveries of the Red Giant 
   Branch Bump(s) in many
   dwarf Spheroidal galaxies of the Local Group. The perspectives for the
   use of this new observational feature to obtain constraints on the Star
   Formation History and Age-Metallicity Relation of nearby galaxies  
   are shortly explored and reviewed. 
   }
   \authorrunning{Michele Bellazzini}
   \titlerunning{The RBG Bump in dSph's}
   \maketitle
%

\section{Introduction}
The so-called Red Giant Branch (RGB) bump is an evolutionary feature predicted
by the theoretical models of the evolution of low-mass stars \citep{iben}.
The H-burning shells moves away from the core of the stars during the evolution
along the RGB. In its outward trip in search of fresh nuclear fuel the
shell encounters the chemical discontinuity left behind by the maximum
penetration of the convective envelope. At this time the luminosity
of the stars drop for a while, until the shell adapts to the new environment,
and then it rises again, burning in a regime of constant H content. As a result
the star passes three times for the same short portion of the RGB evolutionary
path, hence the evolutionary rate significantly slows down in this phase. As a
consequence, at the RGB bump level the stars of a Simple Stellar Population 
(SSP, e.g. a population of coeval and chemically homogeneous stars) pile up,
producing a bump in the Luminosity Function (LF).

%
   \begin{figure*}
   \centering
   \resizebox{\hsize}{!}{\includegraphics[clip=true]{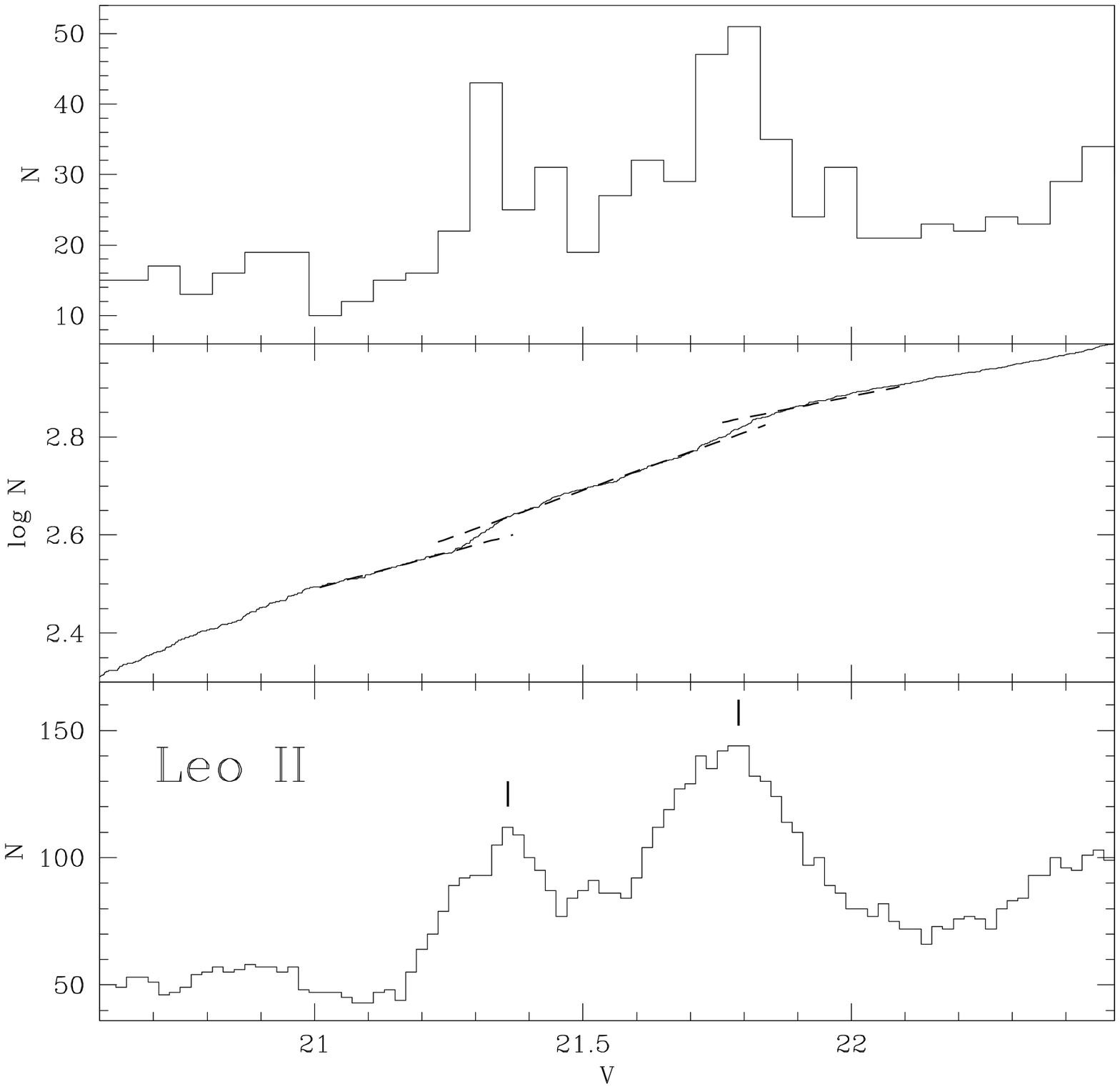}
   \includegraphics[clip=true]{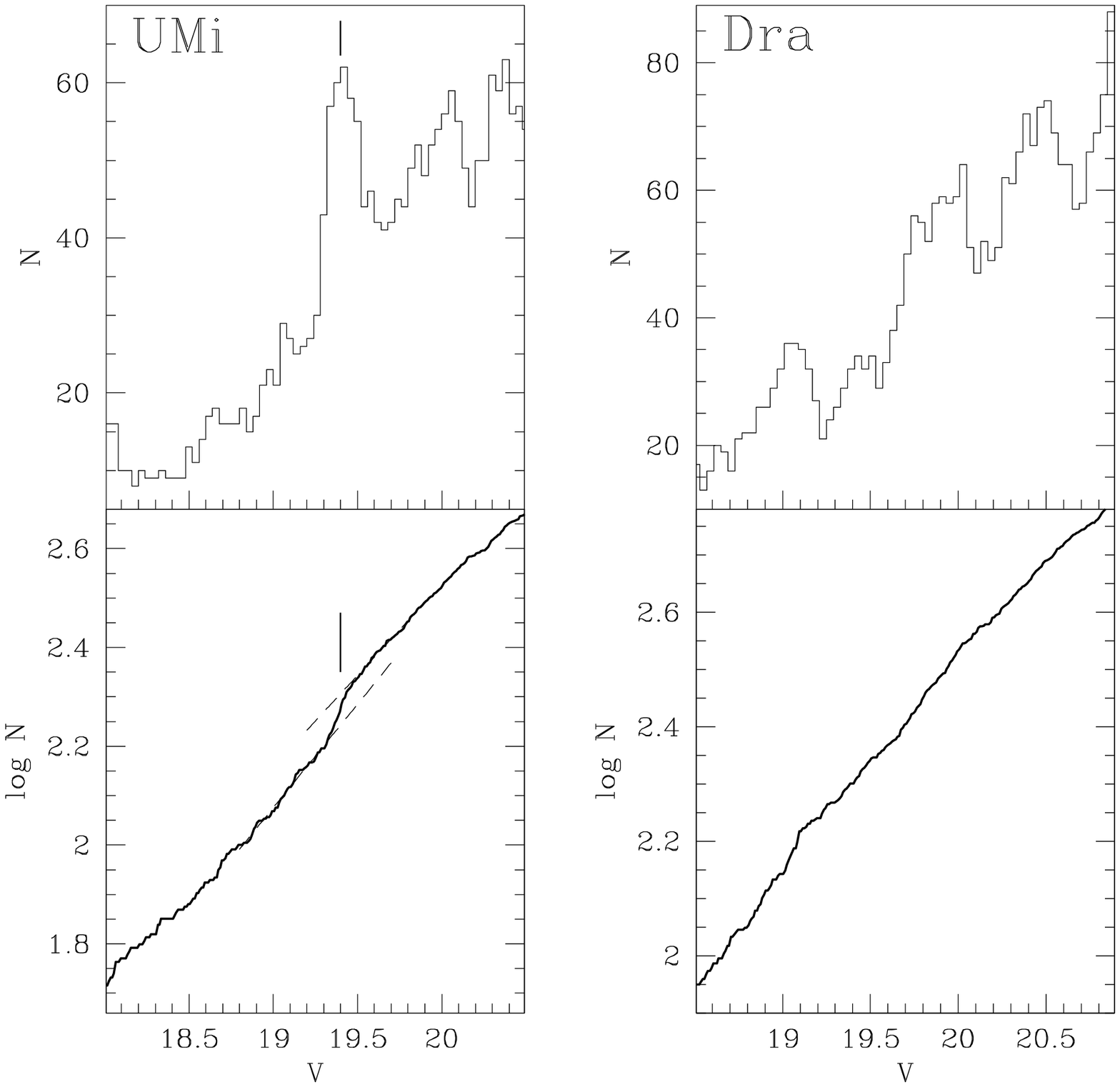}}
     \caption{Left panel: detection of a double RGB bump in Leo II. 
     Central panel: the single RGB bump detected in UMi. Right panel:
     no significant bump is detected in the LF of Draco. The change of slope
     in the cumulative LF is the key indicator for the safe detection of the
     bump \citep{ffp}.
               }  
        \label{F1}
    \end{figure*}
%

From the observational point of view, the first detection of the RGB bump in a
globular cluster (GC) dates back to the '80 \citep{king}. 
At the present time the bump has been observed in several tens of GCs, the
detection technique is quite established \citep{ffp} and the luminosity of the
bump is well predicted by theoretical models 
\cite[see][hereafter F99, and references therein]{zoc,f99}.

Only very recently single and double RGB bumps have been detected in dwarf
spheroidal galaxies (dSph) \citep{sculp,sex,sgr,umidra}, a fact that was somehow
unespected given the nature of composite stellar populations (CSS, e.g., 
containing a mixture of stars of different ages and metallicities) of these 
stellar system.
Here we shortly report on the observational results in this field and provide
some clues on the possible application of this new observable in the study of 
CSS.

\section{The RGB bump in dSph galaxies}

\cite{sculp} reported the detection of two distinct RGB bumps in the Sculptor
dSph. They interpret such double detection, together with the peculiar 
Horizontal Branch (HB) morphology, as evidence for the presence of two main
populations (both quite old) having different metallicities. A very similar
occurrence was found in the Sextans dSph by \cite{sex} and it was interpreted in
an analogous way. In both the above cases the signal of the bumps in the
LF is quite weak and the statistical significance of the observed peaks may be
questioned. In the left panel of Fig.~1 a clear (and strongly significant)
detection of a double bump in the Leo II is presented, as recently obtained by 
our group \citep{leo}. This result proofs beyond any doubt that double bumps do
occur in the RGB-LF of real galaxies (see also Sect. 3, below).

   \begin{figure}
   \centering
   \includegraphics[width=6.5cm]{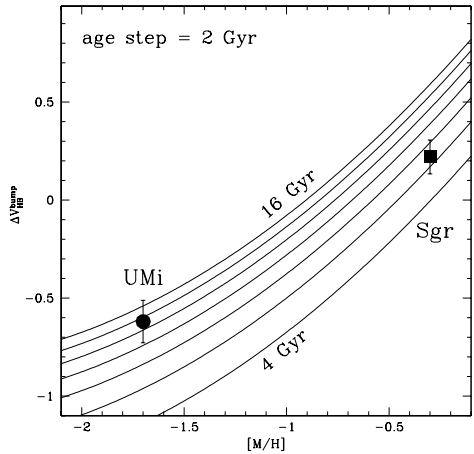}
      \caption{The difference between the magnitude of the bump and that of the
      ZAHB is plotted versus the global metallicity for the UMi and Sgr dSph's.
      The continuous lines are isochrones from Eq. 3 and 4 by F99. Both galaxies
      are dominated by a main population well characterized in metal content and
      age, but the respective SFHs are much different. UMi produced the large
      majority of its stars at early epochs and low metallicity while the main
      star formation episode in Sgr occurred much later, when the chemical
      enrichment of the galaxy was in a quite advanced stage.}
         \label{F2}
   \end{figure}

Single RGB bumps have been clearly detected in the Sgr dSph \citep{sgr} and
in Ursa Minor, while the RGB LF of Draco doesn't show any significant bump 
\cite[][see Fig. 1]{umidra}. 
Thus the present observational scenario provides
cases of {\em no bump}, {\em single} and {\em double bumps}, 
a remarkable variety
considering that the searched sample is presently limited to six dSph
galaxies. 

The luminosity of the RGB bump of a given population depends on its metal 
content and on its age, hence the bumps observed in dSph's may possibly provide
interesting constraints on their Star Formation History (SFH) and Chemical
Enrichment History (CEH). 
We plan to study the behaviour of the RGB bump in composite
population by means of extensive simulations of CSPs under different assumptions
about SFH and CEH. Only such a detailed analysis would provide the necessary
guidelines for a correct interpretation of the bumps observed in CSPs.
Preliminary toy models suggests that a detectable bump may be originated either
by (a) a relatively short ($t_{SF}\le 2$ Gyr) and intense star formation 
episode, or by (b) a relatively long ($t_{SF}\sim 4$ Gyr) period of star 
formation associated with a low rate of chemical enrichment. The latter case is
not surprizing since the luminosity of the bump is much more sensitive to metal
content than age. The above results suggest that the luminosity of the 
RGB bump (as well as  its {\em shape} on the RGB-LF) may ultimately
provide very useful clues on the Age-Metallicity relation of a galaxy, in
particular at old ages. At present, the best exemple of the use of the RGB 
bump as an independent tool to constrain the SFH of a dSph is probably the study
of Sgr by \citet{sgr} (see also Monaco, this meeting). A comparison with the
case of Ursa Minor is presented in Fig.~2.

   \begin{figure}
   \centering
   \includegraphics[width=6.5cm]{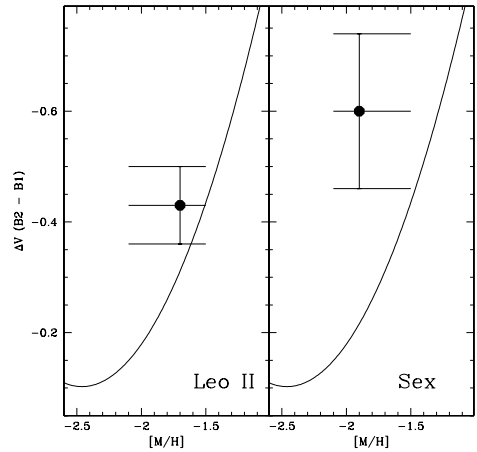}
      \caption{The difference in magnitude between the brigther bump (B2)
      and the fainter bump (B1) for Leo II and Sex is 
      compared to
      the expected difference between the AGB and the RGB bumps as a function of
      metallicity, according to the relations given by F99. The horizontal
      error bars cover the whole range of the possible values of [M/H] for the
      population associated to the main bump (B1).}
         \label{F3}
   \end{figure}

\section{Contamination by the AGB bump?}

Many colleagues, also at this meeting, expressed the doubt that the cases of
{\em double} bumps (at present Scu, Sex and Leo II) may be due to a
misinterpretation of the brighter bump (B2) that may be in fact the AGB bump of
the population that generated the main (fainter) bump (B1)
instead of a second RGB bump. The detailed analysis of this hypothesis is
in progress. Nevertheless there are already serious arguments against this
possibility. First, if this was the case we should have detected a double bump
in any of the galaxies in which we found a RGB bump and this is in contrast with
the observations for UMi and Sgr (note that Sgr is the case with the highest
statistical significance). Second, in observed SSP in which the AGB bump can be
easily identified, the number of stars in the peak of the RGB bump is $\sim 2$
times larger than in the AGB bump. In all the double-bump cases the
ratio of the number of stars in the LF peaks associated to B1 and B2 is 
$\sim 1$. 
This means that the signal (number of stars) expected for the AGB bump 
is much lower than what it is observed in all the B2 bumps detected to date. 
Finally, in Fig.~3 we
compare the observed magnitude difference between B2 and B1 for Leo II and Sex,
with the expected magnitude difference between the AGB and RGB bumps as a
function of metallicity (from Eq.~6 and 7 by F99). In both cases B2 is
significantly brighter than the expected AGB bump and the observed difference
can be (marginally) reconciled with the {\em AGB bump hypothesis} only assuming
rather unlikely average metallicity \citep{shet2}.

Therefore AGB bumps are very unlikely canditates for B2 and double bumps are
probably both bona-fide RGB bumps. Assuming that the B2 bumps are associated
with the most metal poor populations observed in the considered dSph's, the
double bump feature suggest that in these galaxies the SFH at early epochs 
was somehow structured in different main episodes \cite[see][]{sex}.

\section{Conclusions}

The recently discovered RGB bumps of dSph galaxies are a promizing new tool 
for the study of the SFH and the AMR of composite stellar systems. 
These features
are particularly sensitive in the old age regime where the time resolution of
the Main Sequence Turn Off (MSTO) is rather low (and hard to exploit in CSPs), 
therefore they can be nicely complemetary to the traditional age indicators. 

A preliminary exploration of 
the possible use of the bump as an age indicator for GCs whose MSTO 
is out of the reach of the present observational facilities also
provided quite encouraging results.
To exploit the full potentiality
of the bump as an age indicator, accurate estimates of the global metallicity 
(e.g., $[M/H]$ which includes also the contribution of $\alpha$-elements) are
required. Such measures are now within reach of high-resolution spectrometers at
8-meters-class telescopes (e.g., HIRES, FLAMES) also for external galaxies 
and will be the subject of great observational effort in the next few years.

\begin{acknowledgements}
This research is partially supported by the MIUR (Cofin) grant p. 2001028879,
assigned to the project {\em Origin and evolution of Stellar Populations in the
Galactic Spheroid}.
\end{acknowledgements}

\bibliographystyle{aa}

\begin{thebibliography}{}
\bibitem[Bellazzini et al.(2001)]{sex} Bellazzini, M., Ferraro,
         F.R., \& Pancino, E., \mnras, 327, L15	 
\bibitem[Bellazzini et al. (2003a)]{umidra} Bellazzini, M., Ferraro, F.R.,
         Origlia, L., Pancino, E., Monaco, L., \& Oliva, E., 2003a, \aj, 124,
	 3222 	 
\bibitem[Bellazzini et al. (2003b)]{leo} Bellazzini, M., Ferraro, F.R.,
         Gennari, N., 2003b, in preparation	 
\bibitem[Ferraro et al.(1999)]{f99} Ferraro, F.R., Messineo, M., Fusi Pecci,
         F., De Palo, A., Straniero, O., Chieffi, A., \& Limongi, M., 1999,
	 \aj, 118, 1738 (F99)
\bibitem[Fusi Pecci et al.(1990)]{ffp} Fusi Pecci, F., Ferraro, F.R., Crocker,
         D.A., Rood, R.T., \& Buonanno, R., 1990, \aap, 238, 95	 
\bibitem[Iben (1968)]{iben} Iben, I., Jr., 1968, Nature, 220, 143	 
\bibitem[King et al.(1984)]{king} King, C.R., Da Costa, G., \& Demarque, P.,
        1984, \apj, 299, 674	 
\bibitem[Majewski et al.(1999)]{sculp} Majewski, S.R., Siegel, M.H., Patterson,
         R.J., Rood, R.T., 1999, \apj, 520, L33	
\bibitem[Monaco et al.(2002)]{sgr} Monaco, L., Ferraro, F.R., Bellazzini, M., \&
         Pancino, E., 2002, \apj, 578, L47	  
\bibitem[Shetrone et al.(2001)]{shet2} Shetrone, M.D., Cot\'e, P., 
         \& Sargent, P.B., 2001, \apj, 548, 592
\bibitem[Zoccali et al.(1999)]{zoc} Zoccali, M., Cassisi, S., Piotto, G., Bono,
G., \& Salaris, M., 1999, \apj, 518, L49	 
\end{thebibliography}

\end{document}